\documentclass{emulateapj}

\usepackage{graphicx}

\usepackage{epsfig}

\usepackage{natbib}  

\usepackage{epstopdf}

\def\simgt{\lower.5ex\hbox{$\; \buildrel > \over \sim \;$}}

\def\simlt{\lower.5ex\hbox{$\; \buildrel < \over \sim \;$}}

\def\c12{$^{12}$C}

\def\neo20{$^{20}$Ne}

\def\al27{$^{27}$Al}

\def\ne22{$^{22}$Ne}

\def\na23{$^{23}$Na}

\def\mg25{$^{25}$Mg}

\def\mag26{$^{26}$Mg}

\newcommand{\msun}{\ensuremath{\, {M}_\odot}}

\shorttitle{NGC 2419: a benchmark for self--enrichment theories?}

\shortauthors{Ventura et al.}

\begin{document}

\title{Super AGB -- AGB evolution and the chemical inventory in NGC 2419 }

\author{Paolo Ventura}

\affil{INAF -- Osservatorio Astronomico di Roma, via Frascati 33, I-00040 Monteporzio, Italy}

\email{paolo.ventura@oa-roma.inaf.it}

\author{Francesca D'Antona}

\affil{INAF -- Osservatorio Astronomico di Roma, via Frascati 33, I-00040 Monteporzio. Italy}


\author{Marcella Di Criscienzo}

\affil{INAF- Osservatorio Astronomico di Capodimonte, Salita Moiariello 16, I-80131 Napoli (Italy)}


\author{Roberta Carini}

\affil{INAF- Osservatorio Astronomico di Roma, via Frascati 33, I-00040 Monteporzio, Italy}


\author{Annibale D'Ercole}

\affil{INAF- Osservatorio Astronomico di Bologna, via Ranzani 1, I-40127 Bologna (Italy)}


\author{Enrico vesperini}

\affil{Department of Astronomy, Indiana University, Bloomington, USA}


\begin{abstract}

We follow the scenario of formation of second generation stars in globular clusters by matter processed by hot bottom 
burning (HBB) in massive asymptotic giant branch (AGB) stars and super--AGB stars (SAGB). In the cluster NGC 2419 
we assume the presence of an extreme population directly formed from the AGB and SAGB ejecta, so we can 
directly compare the yields for a metallicity Z=0.0003 with the chemical inventory of the cluster NGC~2419.

At such a low metallicity, the HBB temperatures (well above 10$^8$K) allow a very advanced nucleosynthesis. 
Masses $\sim$6\msun deplete Mg and synthesize Si, going beyond Al, so this latter element results only moderately 
enhanced; sodium can not be enhanced. The models are consistent with the observations, although the predicted Mg 
depletion is not as strong as in the observed stars. We predict that the oxygen abundance must be depleted by a 
huge factor ($>$50) in the Mg--poor stars. The HBB temperatures are close to the region where other p--capture 
reactions on heavier nuclei become possible. 
We show that high potassium abundance found in Mg--poor stars can be achieved during HBB, by p--captures on the 
argon nuclei, if the relevant cross section(s) are larger than listed in the literature or if the HBB temperature 
is higher. Finally, we speculate that some calcium production is occurring owing to proton capture on potassium. 
We emphasize the importance of a strong effort to measure a larger sample of abundances in this cluster. 

\end{abstract}


\section{Introduction}
\label{sec:intro}
Photometric and spectroscopic analysis of Globular Cluster (hereafter GC) stars in the last 
decades have changed the traditional framework describing the stellar content of these 
systems: it is now clear that most GCs harbour multiple stellar populations, differing 
in their  original chemistry and characterized by a spread in light elements such as aluminium, 
sodium, oxygen and magnesium \citep[][and references therein]{gratton2012}. These results 
were reinforced by high--quality photometric data, showing that some GCs harbour multiple 
main sequences, that can be interpreted by invoking the presence of a helium--enriched 
population \citep{norris2004, dantona2005, piotto2007, milone2012}. This finding is in 
agreement with the suggestion that helium could be the second parameter traditionally 
invoked to account for the anomalous shape of the Horizontal Branch (HB) of some GCs 
\citep{dantona2002, caloi2005}.

The above results indicate that the stars with anomalous chemistry were born from matter 
contaminated by advanced p--capture nucleosynthesis,  exposed to temperatures sufficiently 
large (exceeding $\sim 100$~MK) to activate the CNO, Ne--Na and Mg--Al nuclear 
channels \citep{prantzos2007}. 
Pollution by ejecta of rapidly rotating massive stars \citep{decressin2007a, decressin2007b}, 
massive binary stars \citep{demink2009}, or intermediate mass Asymptotic Giant Branch (AGB) 
and Super Asymptotic Giant Branch (SAGB)\footnote{SAGB stars are the stars that 
ignite carbon in conditions of partial degeneracy, form an O--Ne core, and then evolve 
through thermal pulses and mass loss like AGB stars \citep[e.g.][]{siess2010}.} stars 
\citep{cottrell, ventura2001} are among the scenarios suggested in the literature. In this 
work, we follow the hydro-dynamical model for the formation of multiple populations by 
\cite{dercole2008} and focus on the hypothesis that AGB and SAGB stars, via the strong 
mass loss at low velocity experienced during their post core Helium burning evolution, 
produced the polluted material, from which the new stars formed.

Notice that Mg depletion, which is an important signature of the extreme anomalies we 
will be dealing with, is a severe constraint on the scenarios, as it can not be found in 
the ejecta of massive stars \citep{decressin2007a}, nor of massive binaries, subject to 
similar nucleosynthesis limitations. Only SAGB models with non extreme mass loss rates  
can predict it \citep{siess2010} or massive AGB models with efficient convection 
\citep[see][]{vcd2012}. We point out that modelling of AGB and SAGB evolution is subject to 
considerable uncertainty concerning the efficiency of the hot bottom burning (HBB) and of 
the third dredge up. Consequently, the AGB--SAGB scenario can explain the chemical patterns 
in GCs only for those model providing compatible yields. Adoption of an efficient convection 
\citep[see the discussion in][]{ventura2005} provides both reasonable sodium yields and 
oxygen depletion (plus some Mg depletion at low metallicity), while models with low 
efficiency of convection and strong dredge up 
\citep{karakas2003, karakas2010, herwig2004, stancliffe2004}
are at variance with AGB--SAGB scenario \citep{fenner2004}.\footnote{Lately, \cite{karakas2012} 
and \cite{lugaro2012} are computing models with more efficient convection, that should be 
able to provide yields closer to the requirements of the GC chemical patterns.} 

\citet{dercole2010, dercole2012cev1} showed that the abundance patterns observed can be
reproduced provided that the star formation of the second generation (SG) starts immediately 
after the epoch of the SNII explosions, and is limited to the first $\sim 100$Myr, thus 
involving only stars of mass $M\geq 5$M$_{\odot}$. The spread observed in the O-Na and 
Mg-Al planes can be explained by dilution of gas ejected by AGBs with pristine gas 
present in the cluster. Of extreme interest is the case of massive GCs, where conditions 
occur for the early formation of a stellar population  directly from the winds of SAGBs, 
with no dilution: these stars are characterized by a strong helium enhancement 
(Y $\simgt 0.35$), and populate the blue MS and the blue tail of the HB of their host clusters.
\cite{dercole2012cev1} showed that the SAGB yields are a very important ingredient of the 
models, for clusters of intermediate metallicity ([Fe/H]$\sim -1.5$, i.e. Z=$10^{-3}$). 
Here we extend our study of the yields from massive AGBs and SAGB stars to the much 
lower metallicity of Z=$3\times 10^{-4}$. This metallicity roughly corresponds to [Fe/H] 
$\sim-2$, the iron content of NGC 2419. This cluster is characterized by a HB hosting a 
blue, faint population, clearly detached from the main component \citep{ripepi2007}, that 
can be explained only by invoking a stellar component greatly enriched in helium
\citep{marcella2011}. Thus, according to the model by \cite{dercole2008}, these stars should 
be born directly from the ejecta of SAGB and from the most massive AGB stars. Consequently, 
the yields computed here should be directly relevant to  the chemical composition of the 
anomalous stars in this clusters. The formation of the SG can be different in other very 
low metallicity clusters such as M~15 \citep{sneden1997}, showing non--extreme HB morphology, 
so here we concentrate on the comparison with NGC~2419 only, and compare our yields to the 
abundances of O, Na, Mg, Al, Si and K recently made available by \cite{cohen2012} and 
\cite{mucciarelli12}. 

We further explore the possibility that the bimodal potassium abundances 
recently discovered by \cite{mucciarelli12}  in NGC~2419 are a signature of the production 
of potassium by proton capture on the argon nuclei in the same HBB environment that produces 
the other abundance anomalies. This can be a powerful indication that the massive AGB/SAGB 
scenario is indeed operating for the formation of multiple populations.

\section{Yields from AGB and SAGB stars}
\label{sec:yields}
The yields from massive AGB stars of interest in this work are determined by Hot Bottom
Burning (HBB), i.e. an advanced, p--capture nucleosynthesis active at the bottom of
the convective envelope \citep{blo2}. \citet{ventura2005} showed that use of the
Full Spectrum of Turbulence \citep[FST,][]{cm1991} model for turbulent convection, 
coupled with the treatment of mass loss by \citet{blo1}, leads easily to HBB conditions 
for all stars with initial mass $M>4M_{\odot}$.

Here we present the yields of AGB and SAGB models experiencing HBB for Z=$3\times 10^{-4}$, 
[$\alpha$/Fe]=+0.4, and put into evidence the difference with the Z=$10^{-3}$ models 
discussed in \citet{vd2009, vd2011}.  A detailed discussion of these models will be presented in a 
forthcoming paper. To allow a more direct comparison with the observations, we show for 
the i--th species the quantity [i/Fe]=$\log(X_i/X_{\rm Fe})-\log(X_i/X_{\rm Fe})_{\odot}$.

\subsection{Oxygen and Sodium}
The degree of the HBB experienced can be deduced by the extent of the oxygen depletion: 
among all the elements considered, oxygen is the only one whose nuclear activity is
a pure destruction process, differently from sodium and aluminium, for which both
creation and destruction channels are active, and magnesium, whose nucleosynthesis
is complicated by the distribution of the overall magnesium content among the three
isotopes \citep{vcd2012}. In the bottom--right panel of Fig.~\ref{figure1} we show the 
typical O--Na pattern for the models from 4 to 7.5\msun. Similar to the Z=$10^{-3}$ 
case, the degree of the p--capture  nucleosynthesis does not increase 
monotonically with mass and reaches a maximum around the threshold of 
$\sim 6M_{\odot}$\footnote{Note that this mass, as also the upper limit of 
7.5-8M$_{\odot}$ above which SNII explosion occurs, are dependent on the assumed overshoot 
from the convective core during the core H--burning phase. In this work we used a moderate 
extra--mixing; if no overshoot was considered, the range of masses involved would 
shift 1-5--2M$_{\odot}$ upwards.} separating the AGB from the SAGB regime. The reason 
\citep[see discussion in][]{vd2011} is in the high mass loss rate of SAGBs, which 
consumes the envelope before a very advanced nucleosynthesis is experienced. In the panel 
we note the straight correlation between the oxygen and sodium content of the ejecta. 
The sodium yields are positive due to the initial phase of sodium synthesis via proton
capture onto $^{22}$Ne, whereas the correlation is due to simultaneous destruction of oxygen and sodium 
once the temperature at which the nucleosynthesis occurs exceeds $\sim 80$MK. The maximum 
depletion in the surface oxygen reaches $[O/Fe] \sim -1.2$ dex at $Z=3\times 10^{-4}$, while it was 
limited to $\sim -0.8$ dex at $Z=10^{-3}$ \citep{vd2009, vd2011}. This is a direct 
consequence of the increasing temperatures at which the bottom of the surface convective 
zone is exposed for models of decreasing metallicity. Notice that in the models with maximum 
oxygen depletion {\sl there can not be any sodium enhancement}. 

To illustrate the dependence of the results on the mass loss treatment, and to compare 
our results with the investigation by \citet{siess2010}, based on smaller rates of mass loss
in the high mass domain, we show, for the 6\msun\ and  7\msun\ models, the results obtained by 
reducing the mass loss rate to 1/4: thanks to the longer time spent in the HBB phase, 
oxygen depletion reaches --1.7~dex for the 6\msun, while sodium is not touched in a significant 
way. A larger O depletion (and further Na depletion) are also found in the 7\msun\ evolution.  
Sodium remains quite large, $[Na/Fe]>+0.3$, only in the mass range 7.2--7.5\msun.

\subsection{Magnesium, Aluminium and Silicon}
\begin{figure}
\center{\resizebox{.48\hsize}{!}{\includegraphics{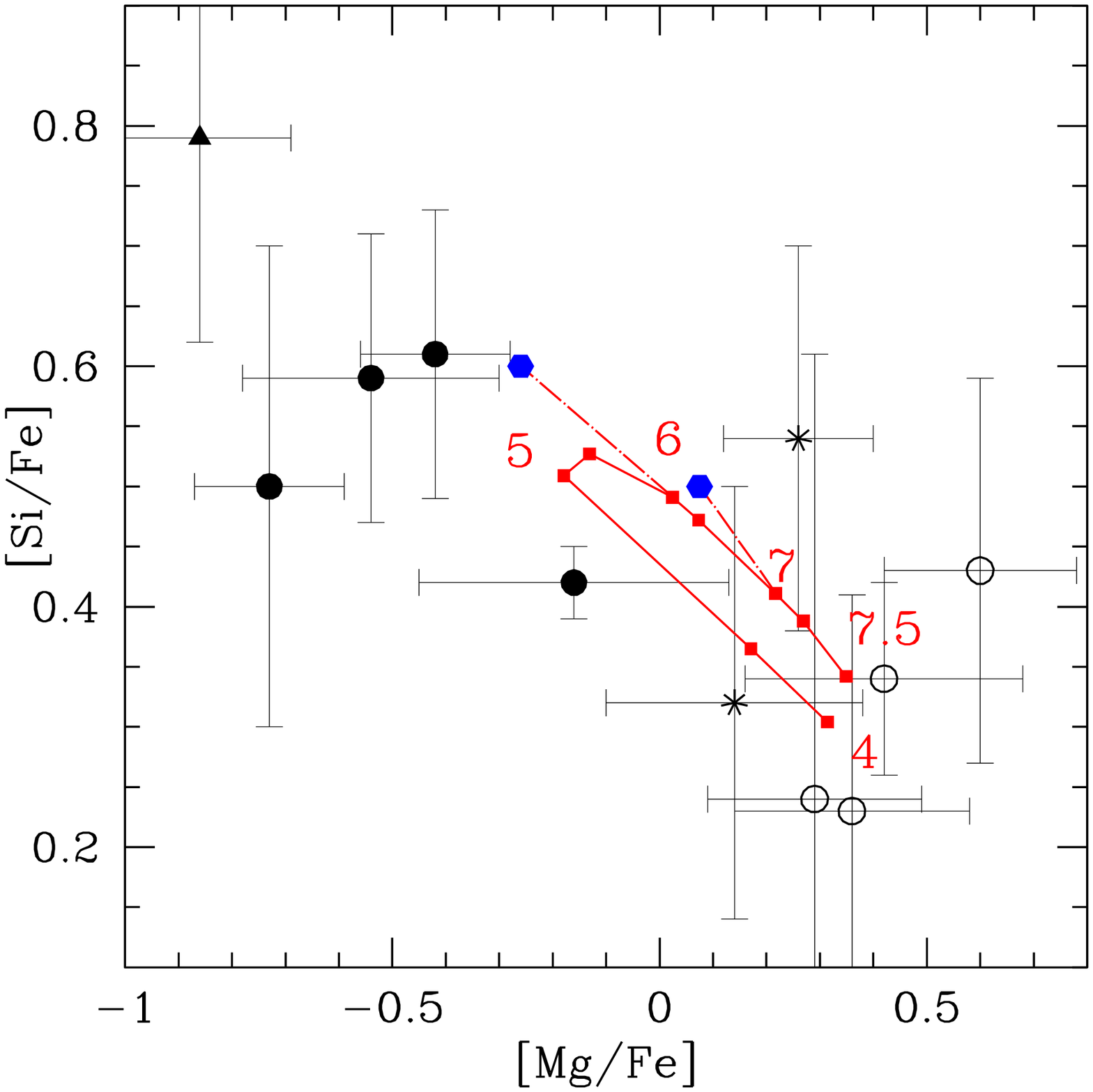}}
\resizebox{.48\hsize}{!}{\includegraphics{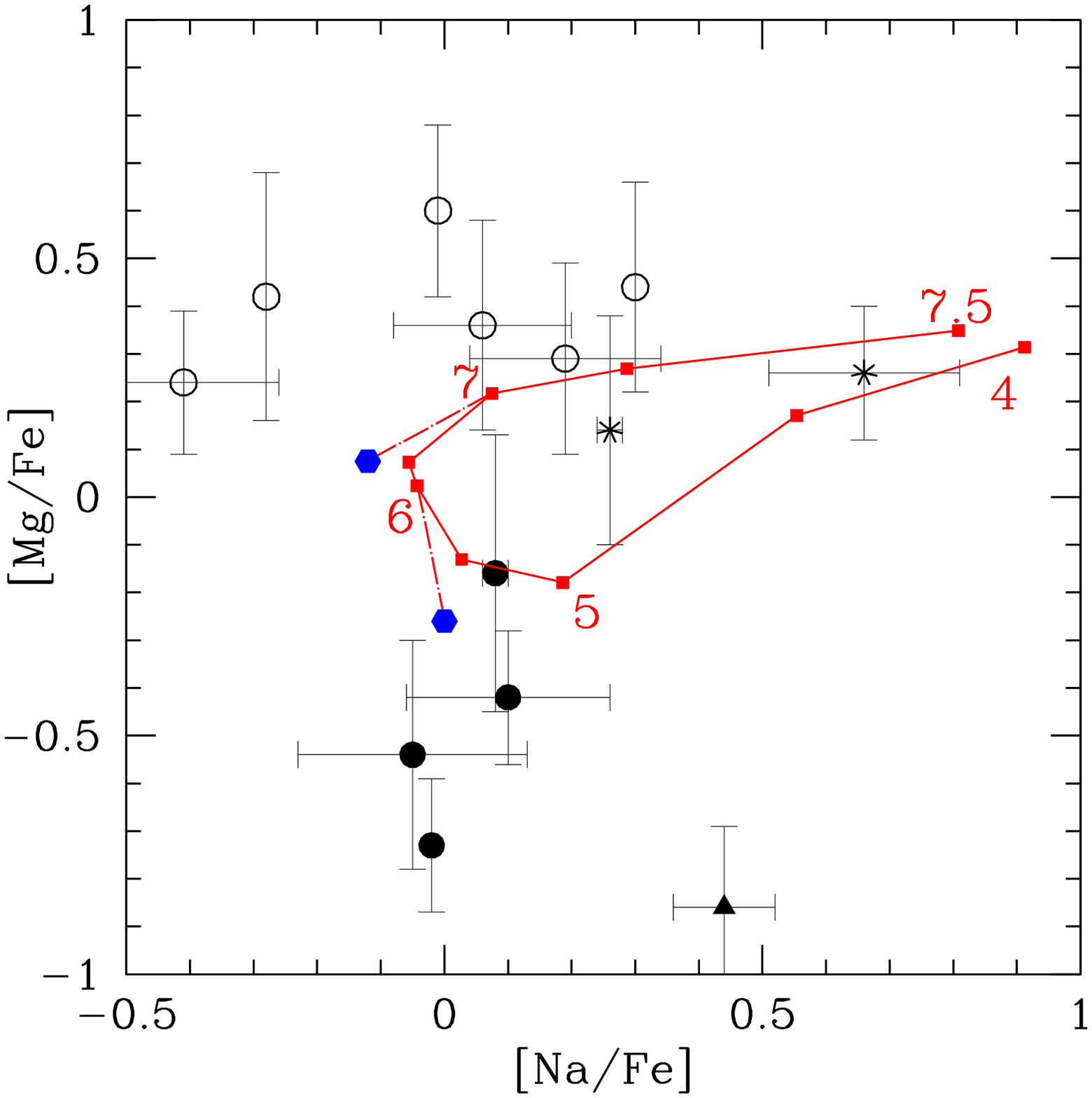}}
\resizebox{.48\hsize}{!}{\includegraphics{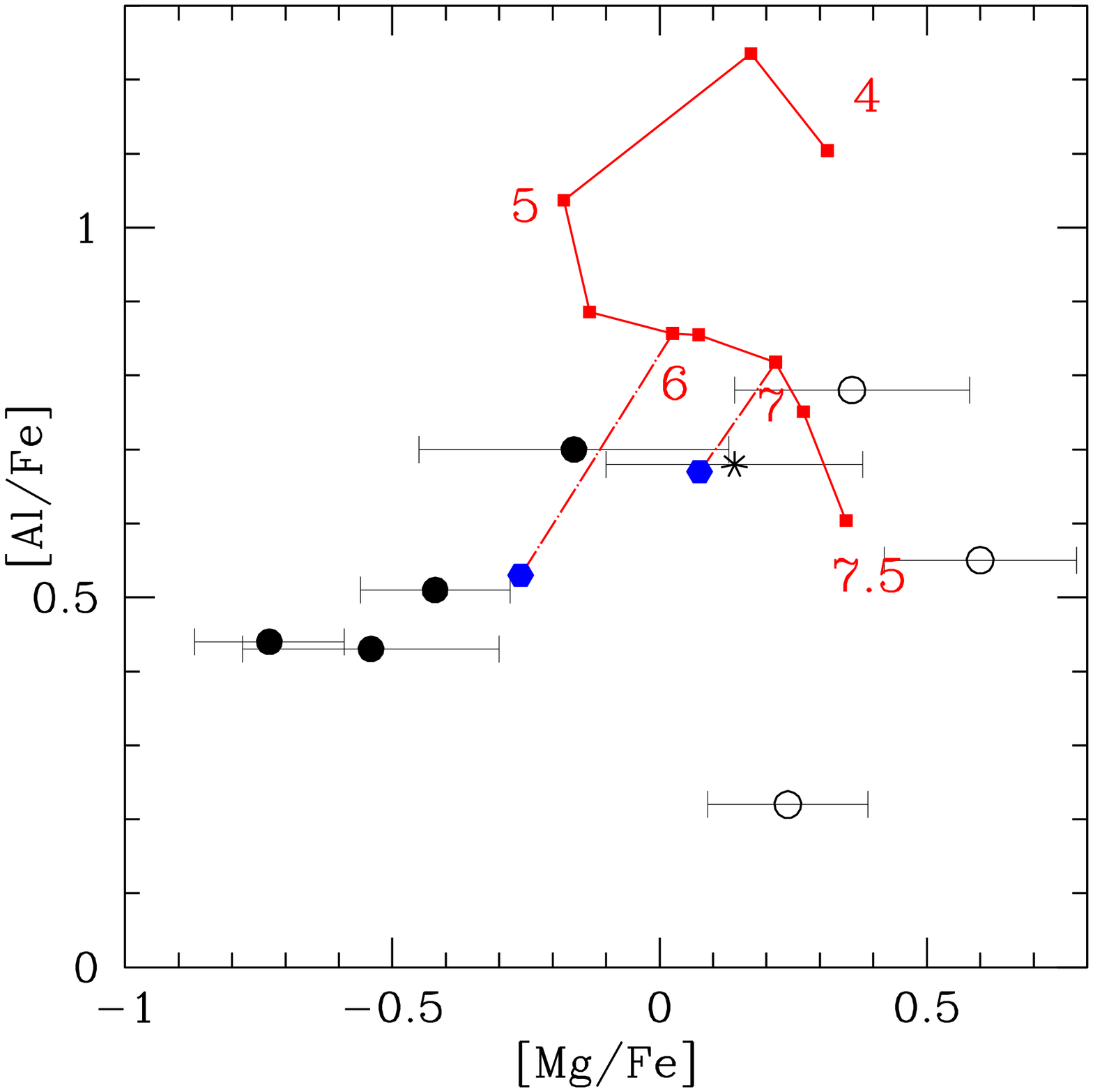}}
\resizebox{.48\hsize}{!}{\includegraphics{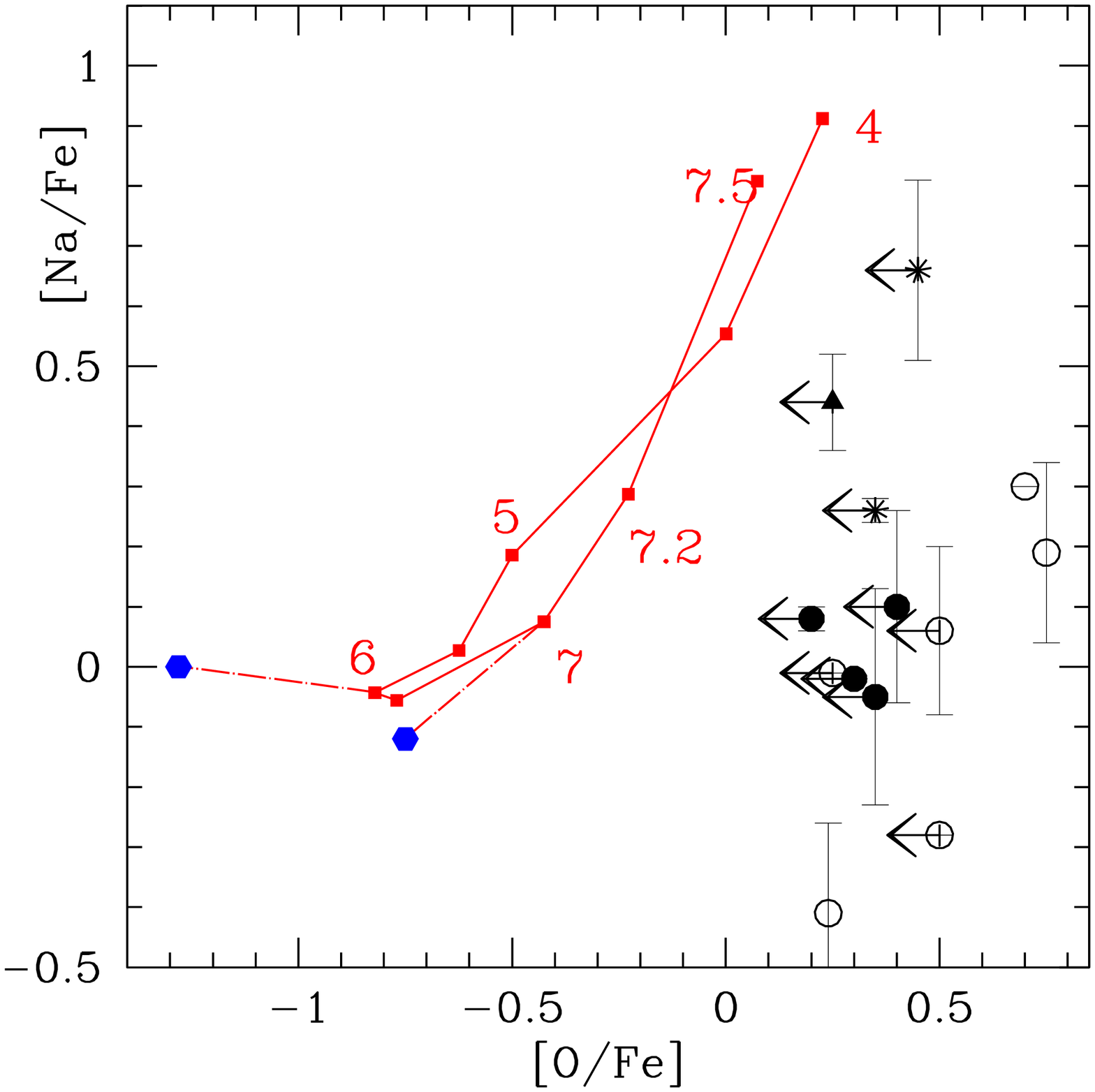}}

}
\vskip +5pt
\caption{Chemical abundances of 13 giants in NGC 2419 (Cohen \& Kirby, 2012). In the 
right--bottom panel, for stars with no abundance determination for oxygen, its upper 
limit is arbitrarily located in the range [O/Fe]=0.25--0.5  in order to show more clearly 
the sodium abundance. The error bars (where provided) are the standard deviations listed 
in Cohen \& Kirby 2012. The data are divided into black dots (Mg--poor giants) and open 
dots (Mg--normal giants). The open dots should represent the first generation stars, 
having typical $\alpha$--rich, Na--normal chemical composition.  The peculiar giant 
Mg--poor and Na--rich is shown as a triangle, the two asterisks denote giants with normal 
Mg and relatively large Na. Black dots, black triangles and asterisks should be compared 
with the theoretical yields for Z=3$\times 10^{-4}$, plotted as  red squares with the 
corresponding mass labelled. The two (blue) hexagons are the yields of  6 and 7\msun, 
when the evolution is computed with mass loss rate reduced to 1/4 of the standard rate. 
\vskip +10pt
}
\label{figure1} 
\end{figure}

The bottom--left panel of Fig.~\ref{figure1} shows the magnesium and aluminium content of 
the ejecta. Although the behavior of the total magnesium is complicated by the equilibria 
of the various p--capture reactions involving the three isotopes, low metallicity models 
can achieve a magnesium destruction larger than that of $Z=10^{-3}$\ models, with the equilibria 
among the various species shifted towards heavier elements. Models of mass
$M>5M_{\odot}$, where the magnesium is most heavily destroyed, show up isotopic ratios 
$^{25}Mg/^{24}Mg\sim 10-30$ and $^{26}Mg/^{24}Mg \sim 1$.
The greater magnesium depletion is not associated with a larger aluminium synthesis, 
because at large temperatures this latter element reaches an asymptotic value of 
[Al/Fe]$\sim +1-+1.1$ dex, at which a balance is reached between the production and 
destruction channels. Models showing the strongest depletion of magnesium show an increase 
of +0.2 dex in the silicon abundance (top--left panel). Reducing the mass loss rate to 
1/4, the Mg depletion in the 6\msun\ increases by -0.1~dex, and Silicon production increases 
by +0.1~dex. In addition, also aluminium decreases. A smaller but similar effect is obtained 
by following the 7\msun\ evolution with reduced mass loss.

\subsection{Helium}
The models associated to the self--enrichment mechanism, with masses exceeding
$\sim 5M_{\odot}$, eject great quantities of helium in the intra--cluster medium, as
a consequence of the second dredge--up, occurring after the core He--burning phase
\citep{vdm2002,pumo2008}. We find a helium mass fraction  Y $\sim$ 0.35--0.37, for masses 
M$\geq 5$\msun.

\section{Comparison with the chemical inventory of NGC~2419}
The large distance of NGC~2419 ($87.5 \pm 3.38$ Kpc from the Sun, \citet{marcella2}) 
has precluded an extensive 
abundance analysis of its stars. However,  some fundamental properties have emerged from 
the work of  \cite{cohen2010,cohen2011} and \cite{cohen2012}, showing the absence of a 
spread in iron, indicating that the system evolved like a typical GC. \cite{cohen2012} 
provide abundances for a total of 13 giants, showing the presence of stars with very low 
Mg abundances, and confirm recent additional data by \cite{mucciarelli12} showing an 
interesting anticorrelation Mg--K. Fig.~\ref{figure1} adds the \cite{cohen2012} data to 
the theoretical patterns. \cite{cohen2012} noticed that the extreme Mg depletion of some 
stars did not correspond to a large increase in the Al abundance. Our models however show 
that the stronger is Mg burning, the more moderate the Al abundance tends to be, as 
nucleosynthesis proceeds towards nuclei of larger atomic number, with Si production. 
This is shown clearly by comparing the standard models to those with reduced mass loss, 
where the time to proceed to more advanced nucleosynthesis is longer, and the Al yields 
are in the range shown by the Mg--poor giants. 

In the panels of Fig.~\ref{figure1} we subdivide Cohen \& Kirby data into two main 
groups: Mg--poor (black dots) and Mg--rich (open dots) giants. We also point out  three 
possibly ``peculiar" giants with moderately large sodium: two giants are shown as asterisks, 
while a third (S1673 in Cohen \& Kirby), very Mg--poor, is shown as a triangle.

First of all, we notice that the oxygen abundances are determined only in a few stars 
having large abundances. We draw the upper limits in the panel O--Na at arbitrary 
locations [O/Fe]=0.25--0.5 to avoid superposition of the points. In this way we show that 
the sodium variation encompasses almost 1~dex, but most points cluster at low 
[Na/Fe]$\simlt 0.3$. The Mg--depleted giants {\sl also show low sodium} (top--right panel): 
these data are compatible with our models showing the strongest HBB processing (5-6\msun). 
S1673 (shown as a triangle) is an exception.

The open points have normal O, Na and Mg, and thus should be plain ``first generation" 
stars; of the two giants denoted as asterisks, one is Mg--normal, Si--normal, so it is 
probably an FG star. The other giant (S1305) is very Na--rich ([Na/Fe]=0.66$\pm$0.15)  
and also Si--rich, but not Mg--poor: we tentatively suggest that it is an SG star formed 
by SAGB ejecta in the 7.2--7.5\msun range, stars that had not enough time to deplete Mg 
too much, but were able to increase Si.

Although the data available are very limited, the observations currently
available suggest that the second generation of  NGC~2419, identified
by \cite{marcella2011} with the blue hook stars can be directly  
formed from the SAGB -- massive AGB ejecta.
Only the Na--rich, Mg--poor giant S1673 is 
out of any qualitative interpretation, as the sodium survival is at variance with the 
strong Mg burning and Si production. We predict that the oxygen abundance in the stars 
with low Mg of NGC~2419 must be extremely low.

\subsection{Understanding K production in the HBB phase}

\begin{figure}
\center{
\resizebox{.95\hsize}{!}{\includegraphics{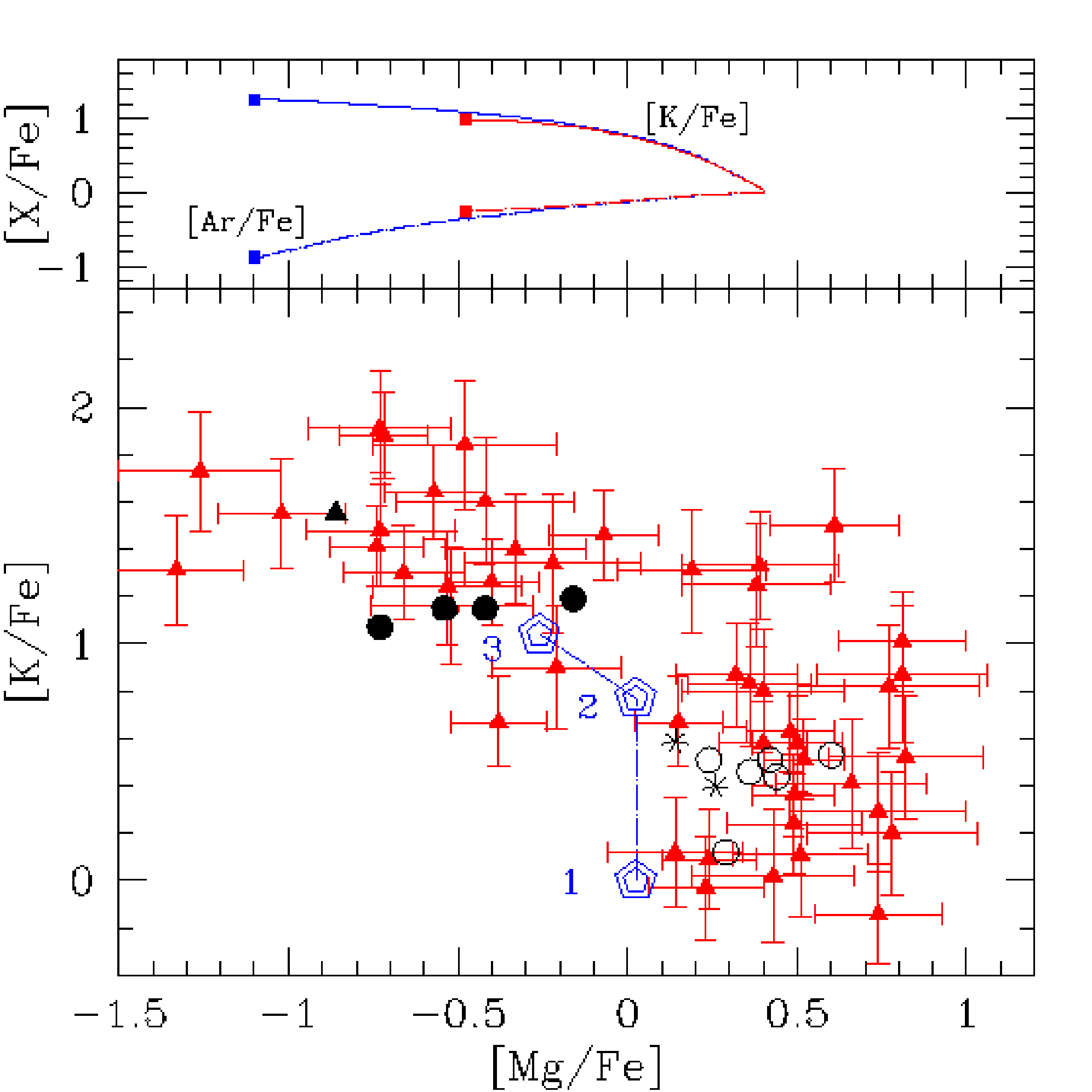}}
}
\vskip +10pt
\caption{The Mg--K data by \cite{mucciarelli12} (red triangles with error bars) and \cite{cohen2012} 
(symbols as in Fig.~1) are compared with the yields obtained by evolving the 6\msun\ star 
in three different cases: 1: with standard $^{38}$Ar(p,$\gamma$)$^{39}$K cross section; 2: 
cross section increased by a factor 100; 3: cross section as in (2), and mass loss rate 
reduced to 1/4. The top panel shows the run of [K/Fe] and [Ar/Fe] versus [Mg/Fe] during the 
evolution in case 2 and 3. We show for each element the logarithm of the ratio of the mass 
fraction to the assumed initial value.
\vskip +10pt}
\label{fmichele} 
\end{figure}

\citet{mucciarelli12} discovered the presence of two distinct, well separated populations 
in NGC~2419, differing in their magnesium and potassium contents, with the Mg--poor 
population significantly enriched in potassium, by a factor $\sim 10$. In Fig.~\ref{fmichele} 
we plot their data together with the data by \cite{cohen2012} that confirm the K--dichotomy. 
Notice that also in this plot the giant pinned up as a (black) triangle is very rich in K, 
confirming the advanced nucleosynthesis (and the difficulty in explaining the high sodium).

Potassium is produced by proton capture on the argon nuclei. The reactions 
involved start from $^{36}$Ar, an $\alpha$--nucleus that will be the most abundant form of 
argon at so low metallicities, as shown by models of galactic chemical evolution
\citep{timmes1995, kobayashi2011}. The chain is
$^{36}$Ar(p,$\gamma$) $^{37}$K(e$^+$,$\nu$) $^{37}$Cl(p,$\gamma$)$^{38}$Ar(p,$\gamma$)$^{39}$K. 
We checked the relevant cross sections by means of the NETGEN tool \citep{aikawa}: all 
of them increase by orders of magnitude at temperature above 10$^8$K, so that the details 
of the envelope structure of our AGB and SAGB models can be relevant to achieve or not the 
required nucleosynthesis. In order to reach the K production, we had to increase the cross 
section $^{38}$Ar(p,$\gamma$)$^{39}$K by a factor 100, but also a stronger HBB, 
leading to temperatures at the base of the envelope exceeding $\sim 150$MK, can achieve 
a similar result. As the initial argon largely exceeds the initial potassium content,  
even a soft activation of the Ar--burning reactions may favor a great increase in the 
potassium abundance. The result of exploratory evolution of the 6\msun\ can be seen in 
the top panel of Fig.~\ref{fmichele}, where the increase in the surface $^{39}$K, and the 
concomitant $^{38}$Ar burning, are shown as a function of [Mg/Fe] for the standard 
track with increased cross section, and for the speculative case in which the mass 
loss rate is reduced to 1/4 of the standard rate. In this latter case, Mg reduction and K production 
are maximized. The main panel of Fig. \ref{fmichele} shows the Mg and K abundances in the 
sample by \cite{mucciarelli12} and in the \cite{cohen2012} sample. The double pentagons 
show the yields of the 6\msun\ evolution, (1) in the standard case (unchanged cross sections), 
(2) in the case of increased cross section, and (3) in the case of increased cross section 
and reduced mass loss. The models show that this path is worth a more complete exploration.

We have discussed so far only the elemental abundances in \cite{cohen2012} for which HBB 
provides a possible nucleosythesis path. Nevertheless, it is necessary to touch the problem of 
Calcium. \cite{cohen2010, cohen2011} found larger Ca abundances in their Mg--poor sample. 
\cite{mucciarelli12} questioned whether the difference in abundance could be attributed 
to the different atmospheric structure, in the presence of such Mg differences, but 
\cite{cohen2012} discuss in depth that this can not be the case. So we must accept the Ca 
variations. Where do these variations come from? At low metallicity, SN ejecta contain more 
$\alpha$--elements than iron \citep{kobayashi2006}, a result dependent on the assumptions 
made on the mass cut \citep[e.g.][]{limongichieffi2003}. A small contamination from SN 
ejecta could then be a viable solution to the calcium problem, although a detailed model 
should be worked out to see whether this is feasible maintaining the lack of iron variations 
and the magnesium depletion. Alternatively, we stretch our results and leave this problem 
open, by asking whether the same p--capture chains could reach $^{40}$Ca from  $^{39}$K.

\section{Conclusions}
The abundances of light elements observed in NGC~2419 have been compared to the chemical 
patterns expected for SAGB and massive AGB evolving through HBB and mass loss, and having 
the low metallicity of this cluster.  We compare directly the yields with observations, 
following the idea that the SG  stars in this cluster are born directly from SAGB plus 
massive AGB ejecta \citep{marcella2011}. The low metallicity of the models allows a very 
strong HBB, especially for masses around $6M_{\odot}$, at the edge between the AGB and 
the SAGB regime. These ejecta are thus predicted to produce the most extreme 
contamination, with a strong depletion of oxygen, a significant reduction of the initial 
magnesium, and only a modest increase in sodium. The models are compatible with the 
abundance trends in NGC~2419. The presence of a K--rich population of Mg--poor stars in 
this cluster is a signature of extreme nucleosynthesis, and we have shown that potassium 
can be produced by proton capture on Argon nuclei, if the relevant cross section is higher 
(by a factor 100) than the standard rate.

The results should be tested by a more detailed spectroscopic investigation. 
Confirming that the Mg--poor stars are extremely depleted in O, and have normal Na would 
support  our model.

\acknowledgements{We thank Marco Limongi for useful discussion. This work has been 
supported by PRIN--INAF 2011: Multiple populations in Globular Clusters: their role in the 
Galaxy assembly (P.I. E. Carretta).E.V. was supported in part by grant NASA-NNX10AD86G.}


\end{document}